\documentclass[10pt,preprint2]{aastex}
\usepackage{wasysym}

\shorttitle{Eccentric Transiting Planets}
\shortauthors{Barnes}

\begin{document}

\title{Effects of Orbital Eccentricity on Extrasolar Planet Transit 
Detection and Lightcurves}
\author{Jason W. Barnes}
\affil{NASA Ames Research Center}
\affil{M/S 244-30}
\affil{Moffett Field, CA  94035}
\email{jason@barnesos.net}

\newpage

\begin{abstract} 

It is shown herein that planets with eccentric orbits are more likely to transit
than circularly orbiting planets with the same semimajor axis by a factor of
$(1-e^2)^{-1}$.  If the orbital parameters of discovered transiting planets are
known, as from follow-up radial velocity observations, then the transit-detected
planet population is easily debiased of this effect.  The duration of a planet's
transit depens upon of its eccentricity and longitude of periastron; transits near
periastron are shorter, and those near apoastron last longer, for a given impact
parameter.  If fitting for the stellar radius with the other transit parameters,
this effect causes a systematic error in the resulting measurements.  If the
stellar radius is instead held fixed at a value measured independently, then it is
possible to place a lower limit on the planet's eccentricity using photometry
alone.  Orbital accelerations cause a difference in the planet's ingress and egress
durations that lead to an asymmetry in the transit lightcurve that could be used
along with the transit velocity measurement to uniquely measure the planet's
eccentricity and longitude of periapsis.  However, the effect is too small to be
measured with current technology.  The habitability of transiting terrestrial
planets found by \emph{Kepler} depends on those planets' orbital eccentricities. 
While \emph{Kepler} will be able to place lower limits on those planets' orbital
eccentricity, the actual value for any given planet will likely remain unknown.

\end{abstract}

\keywords{occultations ---  planets and satellites: individual (HD80606b,
HD147506b) --- techniques: photometric}

\section{INTRODUCTION}

There are presently 21 extrasolar planets known to transit their parent stars ({\tt
http://exoplanet.eu/}).  Radial velocity measurements of all but one of them are
consistent (within errors) with circular orbits; i.e., zero eccentricity
\citep[\emph{e.g.},][]{2005ApJ...629L.121L}.  Presumably, any initial eccentricity
in those orbits has since been damped by tidal circularization
\citep{2000ApJ...537L..61T}.  In light of the discovery of the first transiting
extrasolar planet with an eccentric orbit, HD147506b, I explore the effect that
orbital eccentricity has on transit lightcurves with an eye toward the data to come
from \emph{CoRoT} and \emph{Kepler}.

Since tidal circularization is most effective at short planet-star distances, as
transit search programs extend into longer-period regimes the prospects for
detecting non-circularly orbiting planets grows.  Perhaps not coincidentally, the
first known transiting planet on an eccentric orbit, HD147506b ($e=0.507$), is the
transiting planet with the longest period (5.63 days) \citep{2007arXiv0705.0126B}. 
Recently, \emph{Spitzer} measurements of the relative timing of the secondary
eclipse of GJ436b have confirmed that planet's nonzero orbital eccentricity and
measured it to be $e=0.14\pm.01$ \citep{2007arXiv0707.3809D}.

With the space-based transit searches of \emph{CoRoT} \citep{2003A&A...405.1137B}
and particularly \emph{Kepler} \citep{2005NewAR..49..478B}, hundreds of transiting
planets will be found that will not have been tidally circularized.  Based on the
findings from radial velocity planet searches \citep{2006ApJ...646..505B}, many of
these newly-discovered transiting planets are likely to follow eccentric orbits. 

Orbital eccentricity has several effects on planetary transits.  The timing of the
transit, relative to that of the secondary eclipse, is a strong function of the
planet's orbital eccentricity and longitude of periastron
\citep[\emph{e.g.},][]{2005ApJ...629L.121L,2007AJ....133.1828W}.  Planets with
eccentric orbits, if sufficiently close to their parent stars, can have their
rotations brought into tidal equilibrium at rotation rates greater than their mean
motions \citep{oblateness.2003}; this affects planetary transit lightcurves via
the planet's oblateness \citep{Seager.oblateness,oblateness.2003}.  These effects
and others have been explored in the context of eclipsing binary stars on eccentric
orbits as well \citep[\emph{e.g.}][]{1972ApJ...174..617N}.

In this paper, I explore three additional ways that orbital eccentricity affects
the transits of extrasolar planets.  First, I calculate the increased transit
probability for planets on eccentric orbits.  Next I point out the variability in
transit duration that results from planets moving faster near periastron and
slower near apoastron. The third effect that I explore is the possibility for
asymmetric transit lightcurves induced as the planet's trajectory evolves between
transit ingress and egress.  I then use least-squares fits of synthetic transit
lightcurves to determine whether or not these effects can be used to constrain the
orbital elements of transiting planets from photometry alone.  

Though orbital eccentricity is not inherent to the planet itself, I will sometimes
refer to planets on eccentric orbits as `eccentric planets' for brevity, following
\citep{Lissauer.1997.RotationfromEccentricAccretion}.

\section{EFFECTS}\label{section:effects}

An extrasolar planet on an eccentric orbit has three primary differences
relative to that same planet on a circular orbit with the same semimajor axis. 
The eccentric planet is more likely to transit, and, if it does, then the
transit duration depends upon both the impact parameter $b$ and the orbital true
anomaly $f$, and the
transit lightcurve may be asymmetric due to accelerations during the transit.

\subsection{Transit Probability}\label{section:probability}

Planets on eccentric orbits are more likely to transit than equivalent planets
with the same semimajor axis but circular orbits.  Though these planets spend a
majority of their \emph{time} at greater asterocentric distances than their
semimajor axes, they spend a majority of their \emph{true anomalies} at smaller
asterocentric distances.  The probability for a planet on a circular orbit to
transit was derived by \citet{1984Icar...58..121B} based on the solid angle swept
out by a planet's shadow: \begin{equation}\label{eq:borucki} p~=~\frac{R_*}{r_\mathrm{p}}
\end{equation} where $p$ is the transit probability, $R_*$ is the stellar radius,
and $r_\mathrm{p}$ is the distance between the planet and the star.

\begin{figure} 
\plotone{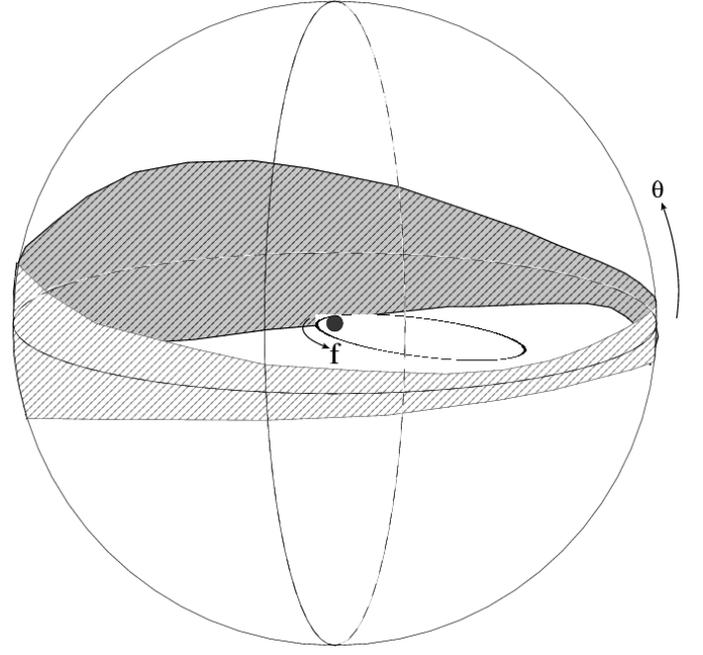}
\caption{Geometry of a planetary orbit and the solid angle swept out by
its shadow, modified from \citet{1984Icar...58..121B} to account for orbital eccentricity.
\label{figure:geometry}}
\end{figure}

Using the method of \citet{1984Icar...58..121B}, then, the transit
probability for an extrasolar planet
is equal to the solid angle swept out by the planet's shadow, a
function of both $f$ and the polar angle from the orbit plane $\theta$, normalized
to $4\pi$ steradians:
\begin{equation}
p~=~\frac{1}{4\pi}\int_0^{2\pi}\int_{\theta_0}^{\theta_1}\cos\theta~\mathrm{d}\theta~\mathrm{d}f
\end{equation}
using $\theta_0$ and $\theta_1$ for the angular
extent of the shadow below and above the orbital plane.  Due to the symmetry of
the problem $\theta_0~=~-\theta_1$.  Geometry sets
$\theta_1~=~\sin^{-1}\frac{R_*}{r_\mathrm{p}}$ (see Figure \ref{figure:geometry}).  
Hence, integrating over $\theta$,
\begin{equation}
p~=~\frac{1}{4\pi}\int_0^{2\pi}~\sin\theta\big|^{\theta_1}_{\theta_0}~\mathrm{d}f
\end{equation}
and plugging in $\theta_0$ and $\theta_1$
\begin{equation}
p~=~\frac{1}{4\pi}\int_0^{2\pi}~\frac{2R_*}{r_\mathrm{p}}~\mathrm{d}f
\end{equation}

For planets on eccentric orbits, $r_\mathrm{p}$ varies with time.  However, the variation
of $r_\mathrm{p}$ as a function of $f$ is all that matters for
determining the solid angle over which the planet will transit.
\begin{equation}\label{eq:orbit}
r_\mathrm{p}~=~\frac{a_\mathrm{p}(1-e^2)}{1+e~\cos f}
\end{equation}
\citep{SolarSystemDynamics} where $e$ is the planet's orbital eccentricity and 
$a_\mathrm{p}$ its semimajor axis.

Plugging in $r_\mathrm{p}$ from Equation \ref{eq:orbit} and integrating over $f$ 
\begin{equation}
p~=~\int_0^{2\pi}{\frac{2R_*~(1+e~\cos f)}{a_\mathrm{p}(1-e^2)}}~\mathrm{d}f  .
\end{equation}
leads to 
\begin{equation}
p~=~\frac{1}{4\pi}~\frac{2R*}{a_\mathrm{p}(1-e^2)}~\int_0^{2\pi}{1+e\cos(f)}~\mathrm{d}f  .
\end{equation}
which leads to the result that for planets on eccentric orbits
\begin{equation}\label{eq:eccentric.probability}
p~=~\frac{R*}{a_\mathrm{p}(1-e^2)}
\end{equation}
since $\int_0^{2\pi}~\mathrm{d}f~=~2\pi$ and 
$\int_0^{2\pi}\cos f~\mathrm{d}f~=~0$.

The above derivation is valid for transit impact parameters $b<=1$.  To exclude
all grazing transits, replace the $R_*$ in the numerator of Equation
\ref{eq:eccentric.probability} with $R_*-R_\mathrm{p}$, where $R_\mathrm{p}$ is
the radius of the transiting planet.  Similarly, to include all transits, no
matter how grazing, the numerator of Equation \ref{eq:eccentric.probability} would
be $R_*+R_\mathrm{p}$.

The increased transit probability for even a planet with a significantly eccentric
orbit with $e=0.5$, similar to that for HD147506b \citep{2007arXiv0705.0126B}, is
modest:  33\%.  However, the increased probability for planets on extremely
eccentric orbits like HD80606b \citep{2001A&A...375L..27N} with $e=0.93$ is
640\%!  Given that 28 of the 224 planets with radial velocity orbits have $e\geq
0.5$ ({\tt http://exoplanet.eu/}), transit surveys should detect a significant
number of planets on eccentric orbits.  Nearly half, 110 out of 224, of radial
velocity planets are more eccentric than our solar system's most eccentric planet,
Mercury ($e=0.2056$).  Planets with extreme orbital eccentricities will be
detected
at a rate decidedly higher than their occurrance would predict given Equation
\ref{eq:borucki}.

This excess will lead to a bias in the raw planet incidence as a function of
semimajor axis derived from \emph{CoRoT} and \emph{Kepler} discoveries.  The
bias can be easily corrected by accounting for the $(1-e^2)^{-1}$ detection
increase factor from Equation \ref{eq:eccentric.probability}, or, more formally,
by using the original \citet{1984Icar...58..121B} probability (Equation
\ref{eq:borucki}) while substituting the instantaneous planet-star distance
at mid-transit for the orbital radius for each detected planet.  

Debiasing requires knowledge of the planet's orbital eccentricity
and longitude of periapsis, which would be difficult to ascertain for planets too
small to induce detectable radial velocity variations (Section
\ref{section:detectability}).  Although this extra step adds complexity, it is
heartening to note that the orbital eccentricity bias induced in planet
distrubutions as derived using the transit method is precisely calculable and
removable.  The orbital eccentricity bias in radial velocity planet surveys (as
results from data gaps near periastron passage, for instance) is known less
precisely and is therefore more challenging to remove.

The differing transit probability at periastron and apoastron can lead to planets
that transit, but have no secondary eclipse if the planet is sufficiently
eccentric, inclined, and transits near periastron.  Alternately there can exist
planets that show secondary eclipses but no primary transit.  In the case where
giant planets have just secondary eclipses, the small secondary eclipse depth
could be mistaken for the primary transit of a terrestrial-sized planet.  Careful
monitoring of reflected-light phase variability (as described in
\citet{2002ApJ...575..493J}) may eliminate this source of systematic error. 
Secondary-eclipse-only planets would need to be very near their parent stars
during secondary eclipse in order to have sufficient reflected light so as to
mimic the transit of a terrestrial-sized planet.  Faint secondary stars in similar
orbits could mimic the terrestrial planet transit regardless of their distance
from the primary star at secondary eclipse.

\subsection{Transit Duration}\label{section:duration}

For a given star mass, planets with the same orbital semimajor axes  have the same
energy per unit mass ($-GM_*/2a_p$); those with higher  eccentricities have lower
specific angular momenta.  As such, those  more highly eccentric objects move
faster near periastron, and slower near  apoastron.  If such an eccentric planet
were to transit, it would then have transits of shorter  or longer duration than
the equivalent planet on a circular orbit transiting with the same impact
parameter ($b$).

\begin{figure} 
\plotone{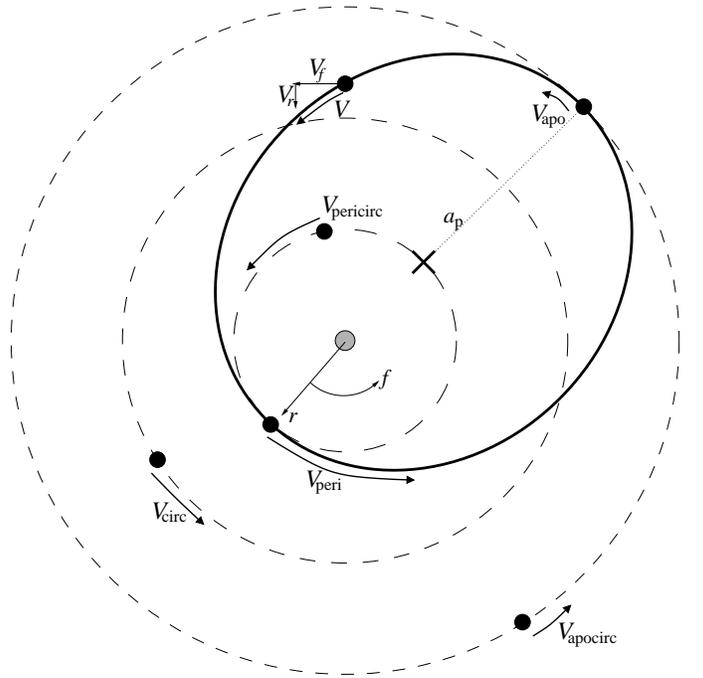}
\caption{This figure illustrates the variables described in Section
\ref{section:duration} of the text, as applied to a hypothetical planet
with $e~=~0.5$.
\label{figure:eccorbit}}
\end{figure}

The velocity of a circularly orbiting planet $V_\mathrm{circ}$ is 
\begin{equation}
V_\mathrm{circ}~=~\sqrt{\frac{GM_*}{a_p}}~. 
\end{equation}
Using conservation of energy, the 
periapsis velocity $V_\mathrm{peri}$ of a planet with orbital eccentricity 
$e$ can be shown to be 
\begin{equation}\label{eq:periastron}
V_\mathrm{peri}~=~\sqrt{\frac{1+e}{1-e}}\sqrt{\frac{GM_*}{a_p}}~=~
\sqrt{\frac{1+e}{1-e}}~V_\mathrm{circ}~=~\sqrt{1+e}~V_\mathrm{pericirc}~.
\end{equation}
In the case of planets discovered by their transits, $a_p$ is set by the 
planet's known orbital period, and hence the comparison to 
$V_\mathrm{circ}$ is most relevant.
For convenience I also compare $V_\mathrm{peri}$ to 
$V_\mathrm{pericirc}$, the velocity of 
a planet orbiting circularly at the eccentric planet's periastron.  
Similarly, at apoapsis 
\begin{equation}\label{eq:apoastron}
V_\mathrm{apo}~=~\sqrt{\frac{1-e}{1+e}}\sqrt{\frac{GM_*}{a_p}}~=~
\sqrt{\frac{1-e}{1+e}}~V_\mathrm{circ}~=~\sqrt{1-e}~V_\mathrm{apocirc} 
\end{equation}
for similarly named variables (see Figure \ref{figure:eccorbit}).  

The pericenter and apocenter velocities behave as expected in their 
extremes.  $V_\mathrm{peri}$ approaches escape velocity 
($\sqrt{2}~V_\mathrm{pericirc}$) as $e\to1$ but is undefined for non-closed 
orbits where $e\geq1$.  $V_\mathrm{apo}\to0$ as $e\to1$, in absolute terms, as a
function of $V_\mathrm{circ}$, and as a function of $V_\mathrm{apocirc}$.

A significantly eccentric planet with $e=0.5$ (similar to HD147506b for which
$e=0.507$) travels  $\sqrt{3}$ ($\sim1.73$) times faster at periapsis than you
would expect given its  period and assuming a circular orbit, and $\sqrt{3}$ times
more slowly at  apoapsis.  This planet's transit duration, if it were to transit
at periastron, would be 58\% as long for a particular impact parameter as the 
transit of its circularly orbiting equivalent.  If transiting at apoastron  (three
times less probable than a periastron transit; see Section 
\ref{section:probability}), the transit would last 73\% longer than the 
equivalent circularly orbiting planet.

A planet on an extremely eccentric orbit like HD80606b ($e=0.93$)
\citep{2001A&A...375L..27N} would have a periastron transit duration only 19\% that of
its same-semimajor-axis circular equivalent, and 72\% that of a planet orbiting
circularly at HD80606b's periastron.  Conversely if HD80606b were to transit near
apoastron, such a transit would last 5.25 times longer than if HD80606b were to
transit in a circular orbit with the same semimajor axis, and 3.78 times longer
than if HD80606b were to transit in a circular orbit at its true apoastron.

The duration of a exoplanetary transit depends on the chord length of the planet's
apparent passage in front of the star ($2R_*\cos b$) and the planet's azimuthal
velocity $V_{f}~=~r\dot{f}$.  In the cases discussed above, when a planet is at
periastron and apoastron, $V_\mathrm{peri}$ and $V_\mathrm{apo}$ are equal to
$V_f$.  The rest of the time, $V_f$ if not equal to the planet's full velocity as
the planet will also have a velocity component radial to the star ($V_r$).  
According to
\citet{SolarSystemDynamics}, $V_f$ varies sinusoidally with $f$:
\begin{equation} \label{eq:velocity}
V_f~=~V_\mathrm{circ}\frac{1+e\cos f}{\sqrt{1-e^2}}~\mathrm{.}
\end{equation}
Hence the planet's velocity is greater than the equivalent circular orbit velocity
for more than half of the orbit as a function of the true anomaly.

\subsection{Transit Symmetry}\label{section:asymmetry}

The dependence of the planet's azimuthal velocity on $f$ as shown in Equation
\ref{eq:velocity} belies to another effect that eccentric orbits have on
transits.  Because $V_f$ changes slightly between ingress and egress (unless the
planet is at periastron or apoastron mid-transit), eccentric planets can produce
asymmetric transit lightcurves.  If a planet transits after periastron and before
apoastron, the time that it takes for the planet to ingress across the star's limb
will be shorter than the time that it takes to egress.  Similarly if the planet is
between apoastron and periastron, then the ingress will be longer than the egress.

\begin{figure} 
\plotone{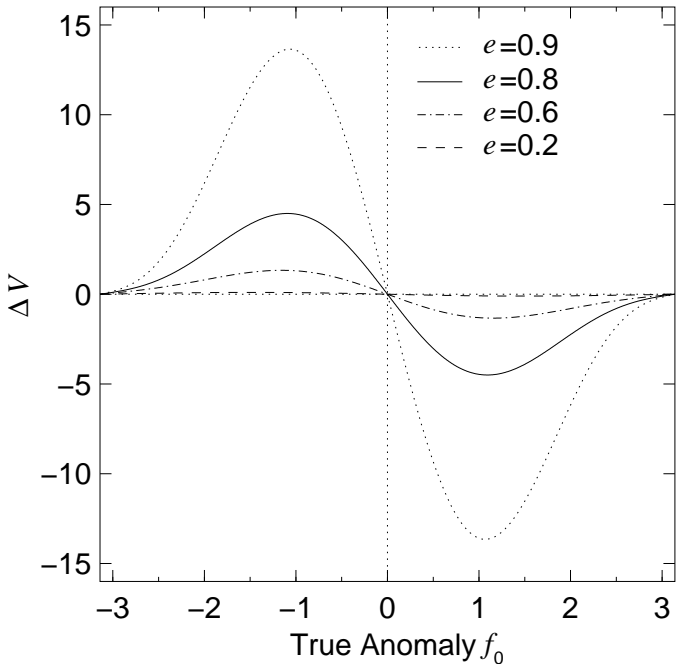}
\caption{Velocity difference between egress and ingress as a function of the
planet's orbital true anomaly at mid-transit, $f_0$.  I have only plotted that
portion of Equation \ref{eq:deltaV} that is a function of $f$ and $e$; to convert to
true $\Delta V$, multiply by $\frac{2R_*V_\mathrm{circ}}{a}$.  As an example, the
multiplier for a planet in an $a_p=0.1$AU orbit around a $1M_\odot$ star is 8.76
km/s.
\label{figure:deltaV}}
\end{figure}

To calculate the velocity difference between ingress and egress, I first take
$V_f$ from Equation \ref{eq:velocity} and differentiate it with respect to $f$:
\begin{equation}
\frac{\mathrm{d}V_f}{\mathrm{d}f}~=~-\frac{e~V_\mathrm{circ}}{\sqrt{1-e^2}}~\sin f \mathrm{.}
\end{equation}
The total ingress-egress velocity difference, $\Delta V$, is equal to
$\frac{\mathrm{d}V_f}{\mathrm{d}f}$ times $\Delta f$, the difference of the true
anomaly between ingress and egress, under the assumption that 
$\frac{\mathrm{d}V_f}{\mathrm{d}f}$ varies negligibly across the transit.  Taking
$f_0$ to be the true anomaly of the planet at mid-transit,
\begin{equation}
\Delta f~=~\frac{2R_*}{r_\mathrm{p}(f_0)}
\end{equation}
assuming that $r_\mathrm{p}$ varies only slowly during the transit.  Now, plugging in $r_\mathrm{p}$
from Equation \ref{eq:orbit},
\begin{equation}\label{eq:deltaV}
\Delta V~=~\frac{\mathrm{d}V_f}{\mathrm{d}f}~\Delta f~=~-
\frac{2R_*eV_\mathrm{circ}}{a(1-e^2)^{3/2}}\sin f_0~(1+e\cos f_0)~\mathrm{.}
\end{equation}
I show a plot of the varying component of $\Delta V$ for various values of $e$ in
Figure \ref{figure:deltaV}.
The \emph{fractional} difference in velocity, $\Delta V/V$, is a bit simpler:
\begin{equation}
\frac{\Delta V}{V}~=~-\frac{2R_*e}{a(1-e^2)}\sin f_0~\mathrm{,}
\end{equation}
and has an evident maximum where $f_0=\pm\pi/2$.  Hence, the greatest
\emph{fractional} variation in ingress and egress duration will occur ninety degrees 
away from periastron and apoastron, as measured in the planet's true anomaly.

\begin{figure} 
\plotone{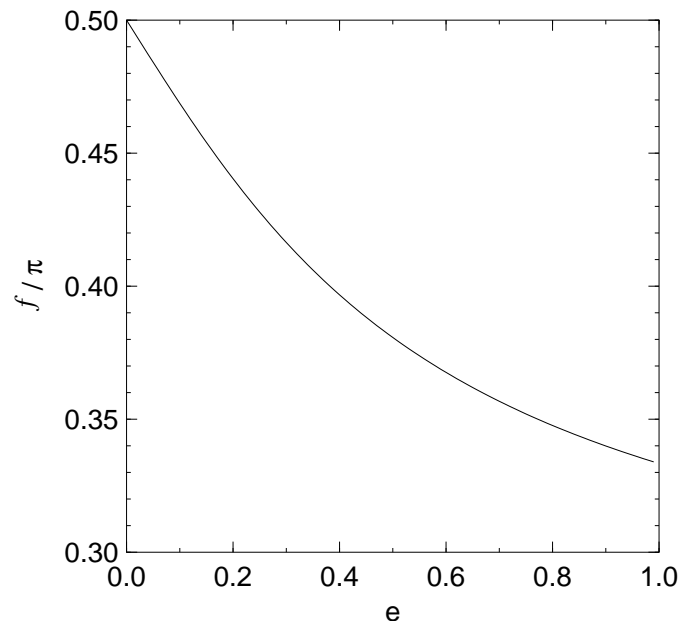}
\caption{The true anomaly at mid-transit 
($f_0$, here in units of radians divided by $\pi$) for which
the difference between a planet's transit ingress and egress velocity is maximized
as a function of the planet's orbital eccentricity, $e$.
\label{figure:maxdeltaV}}
\end{figure}

To determine where the \emph{absolute} velocity difference $\Delta V$ is maximized, I
differentiate $\Delta V$ with respect to $f$ and set the result equal to zero:
\begin{equation}
0~=~\frac{\mathrm{d}\Delta
V}{\mathrm{d}f}~=~-\frac{2R_*eV_\mathrm{circ}}{a(1-e^2)^\frac{3}{2}}
\frac{\mathrm{d}\big[\sin f_0+e\sin f_0\cos f_0\big]}{\mathrm{d}f}~\mathrm{.}
\end{equation}
Differentiating and using the double-angle formula $\sin (2x)=2\sin x \cos x$, I
determine that 
\begin{equation}\label{eq:maxdeltaV}
\cos f_0+e\cos (2f_0)~=~0~\mathrm{.}
\end{equation}
I show the solution to this equation in Figure \ref{figure:maxdeltaV}. 
The maximum velocity difference occurs at $f_0=\pm\pi/2$ for infinitessimal 
eccentricities, and approaches $f_0=\pm\pi/3$ as $e\to1$.

\begin{figure} 
\plotone{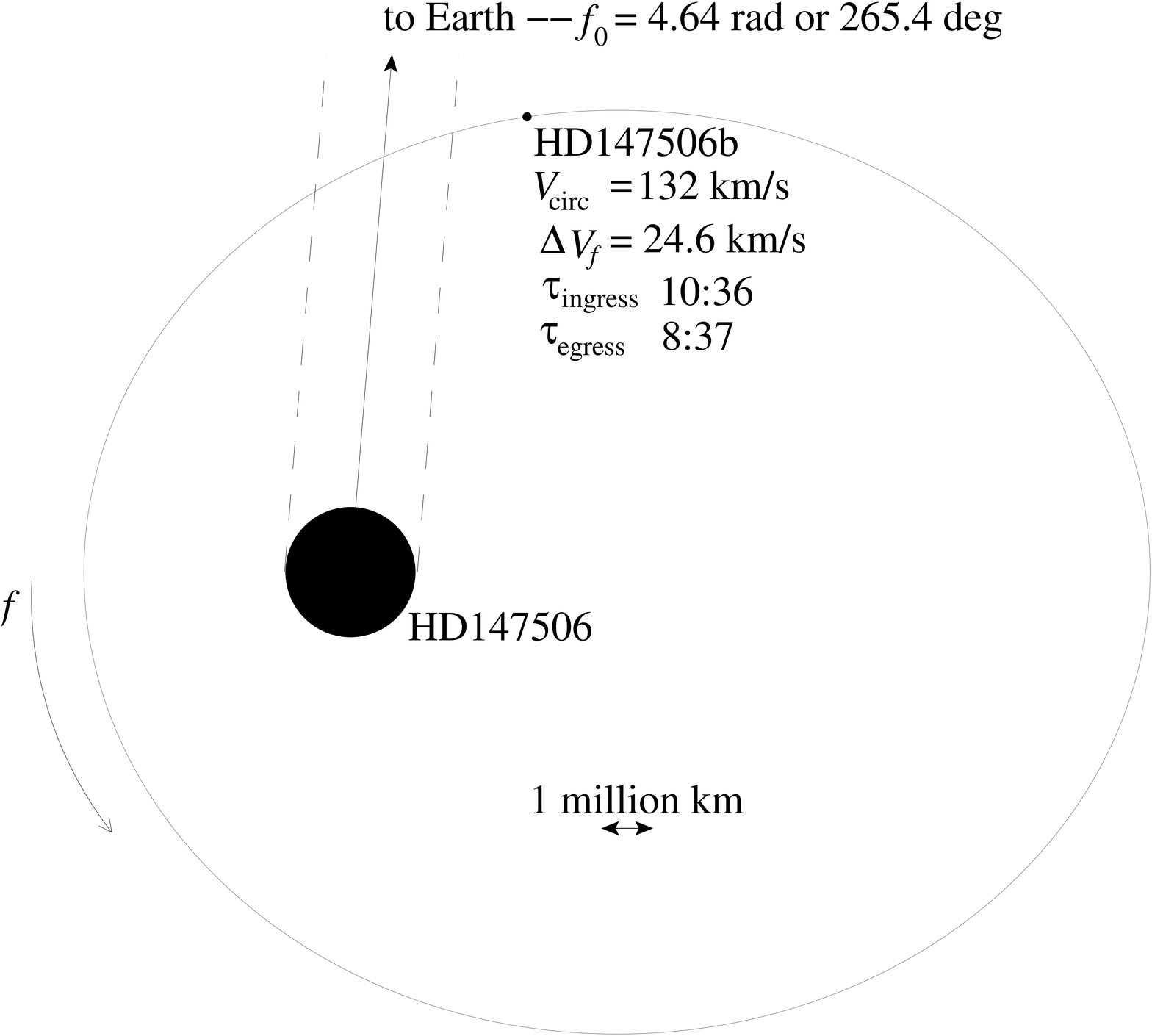}
\caption{To-scale diagram of the HD147506 system.  Ingress and egress times listed
are in minutes~:~seconds format.  Figure concept inspired by Gregory Laughlin's
{\tt http://oklo.org/} website.
\label{figure:HD147506b}}
\end{figure}

To illustrate the consequences of this effect, I apply the results derived above
to newly-discovered eccentric transiting planet HD147506b
\citep{2007arXiv0705.0126B}.  HD147506b was determined by its discoverers,
\citet{2007arXiv0705.0126B}, to have a radius of $1.18~\mathrm{R_{Jup}}$, an
orbital semimajor axis of $0.0685~\mathrm{AU}$, an orbital eccentricity of
$0.507$, and a longitude of periastron of $184.6^\circ$.  The planet's parent
star, HD147506, was determined to have a radius of $1.8~\mathrm{R_\odot}$ and a
mass of $1.35~\mathrm{M_\odot}$ (note that refined system parameters were
published by \citet{2007arXiv0707.0679L} while this paper was in review -- the new
values do not change the qualitative results that I describe here, but future work
should employ these newer values).  I show a to-scale schematic of the system in
Figure \ref{figure:HD147506b}.

From Equation \ref{eq:velocity} HD147506b's $V_\mathrm{circ}$ is
$132.2~\mathrm{km~s^{-1}}$.  This planet's orbital parameters are particularly
favorable with respect to maximizing the magnitude of $\Delta~V_f$.  HD147506b's
large eccentricity, fast $V_\mathrm{circ}$, and nearly ideal true anomaly at
midtransit ($f_\mathrm{0}=4.84$, very close to $-\pi/2$), combined with HD147506's
large stellar radius yield $\Delta~V_f~=~-24.6~\mathrm{km~s^{-1}}$.  Calculating
the ingress and egress times $\tau$ using
\begin{equation}
\tau = \frac{R_\mathrm{p}}{V_f~\cos(\sin^{-1}b)}
\end{equation}
with the appropriate $V_f$ values for the transit ingress and egress of HD147506b
yields $\tau_\mathrm{ingress}~=~10~\mathrm{min}~36~\mathrm{sec}$ and
$\tau_\mathrm{egress}~=~8~\mathrm{min}~37~\mathrm{sec}$.

\section{DETECTABILITY}\label{section:detectability}

The \emph{CoRoT} and \emph{Kepler} missions will discover hundreds of new
transiting extrasolar planets \citep{2003A&A...405.1137B,2005NewAR..49..478B}. 
The most massive of these will be amenable to radial velocity follow-up
observations to measure their masses; a more complete set of radial velocity
measurements (\emph{i.e.}, covering the full orbit and not just the radial
velocity maxima and minima, which are all that would be required to determine mass
assuming a circular orbit and the epoch and period as established by the transit)
can determine the planets' orbital eccentricities and longitudes of periastron. 
However for planets of Neptune-mass and smaller, in larger orbits, or orbiting
fainter stars, radial velocity follow-up will not be practical or in some cases
possible.   In those cases, it would be useful to attempt to constrain the orbital
eccentricity of those planets using transit photometry alone.  Such a
determination would bear critically on the climatic variability of Earth-like
worlds; highly eccentric planets may not be habitable even if their orbital
semimajor axes place them within their stars' habitable zone.

To determine the photometric detectability of the transit duration and asymmetry
effects of eccentric planet orbits, I create synthetic transit lightcurves that I
then fit as if the orbit were circular.  I assume knowledge of the planet's period,
as that value will be measured by the time between transits in \emph{Kepler} and
\emph{CoRoT} data.  I calculate both the synthetic lightcurves and the best-fit
solutions using the analytical approximation of \citet{2002ApJ...580L.171M}.  The
\citet{2002ApJ...580L.171M} formulation assumes that the portion of the star
covered by the planet has uniform surface brightness, but that brightness accounts
for limb-darkening; hence it is least accurate for ingress and egress.  However,
the detrimental effects on this particular calculation should be minimal since both
the synthetic and best-fit lightcurves should show the same systematic errors,
which, when subtracted, should leave a good estimate for the proper fit residuals. 
The effects of light-travel delay illustrated by \citet{2005ApJ...623L..45L} are
not included.  

For larger planets, those where the ingress and egress are temporally resolved,
that have high signal-to-noise lightcurves, I first fit for $R_*$, $R_\mathrm{p}$,
the transit impact parameter $b$, and a stellar limb darkening coefficient $c_1$,
the treatment applied to HD209458b by \citet{2001ApJ...552..699B}.  I assumed a
$1~R_\mathrm{Jup}$ planet orbiting a $1~R_\odot$, $1~M_\odot$ star with
$a_p=0.1~\mathrm{AU}$ and $e=0.5$.  Since $R_{\mathrm{p}}/R_*$ is set by the
transit depth, and $b$ by the ingress/egress time relative to the total transit
duration, orbital eccentricity in this type of fit systematically affects $R_*$,
which governs the total transit timescale.  In my test runs, the best-fit
$R_{*\mathrm{measured}}$ varies as $R_{*\mathrm{measured}}=R_* V_\mathrm{circ}
V_{f_0}^{-1}$.  Hence assuming a circular orbit when the orbit is actually
eccentric leads to a systematic error in the measurement of $R_*$ and, by
extension, $R_\mathrm{p}$ ($R_\mathrm{p}/R_*$ is unaffected).  The measured values
are smaller than the actual values if the planet is near periastron, and are
larger if the planet is near apoastron.

This systematic error can be addressed by assuming a stellar radius, as could be
estimated by other observations such as parallax, spectral type, and stellar
absolute magnitude.  In this case, the transit timescale can be set by fitting
explicitly for $V_{f_0}$ in addition to $r_\mathrm{p}$, $b$, and $c_1$.  However, without
knowledge of the mid-transit true anomaly $f_0$, $V_{f_0}$ cannot uniquely
determine the orbital eccentricity.  Instead, $V_{f_0}$ can \emph{constrain} $e$
if we allow that the planet must have a minimum eccentricity in order that $V_f$
reach $V_{f_0}$.  If $V_{f_0}~>~V_\mathrm{circ}$, then from Equation
\ref{eq:periastron}
\begin{equation}\label{eq:eminperi}
e~\geq~\frac{\big(\frac{V_{f_0}}{V_\mathrm{circ}}\big)^2-1}
         {\big(\frac{V_{f_0}}{V_\mathrm{circ}}\big)^2+1}
\end{equation}
and if $V_{f_0}~<~V_\mathrm{circ}$, then similarly from Equation
\ref{eq:apoastron}
\begin{equation}\label{eq:eminapo}
e~\geq~\frac{1-\big(\frac{V_{f_0}}{V_\mathrm{circ}}\big)^2}
         {1+\big(\frac{V_{f_0}}{V_\mathrm{circ}}\big)^2}~\mathrm{.}
\end{equation}
The above lower limits can only be placed when the planet's ingress and egress are
resolved.  In order to constrain the eccentricity of terrestrial-sized planets,
both high temporal cadence and high photometric precision would be necessary. 
Resolving the ingress of the Earth, with $\tau_\mathrm{ingress}=7.04~\mathrm{minutes}$ at $b=0$,
will not be possible given \emph{Kepler} data alone.  Coaddition of
multiple transits from multiple telescopes might help to constrain the ingress 
times for detected transiting terrestrial planets.

\begin{figure} 
\plotone{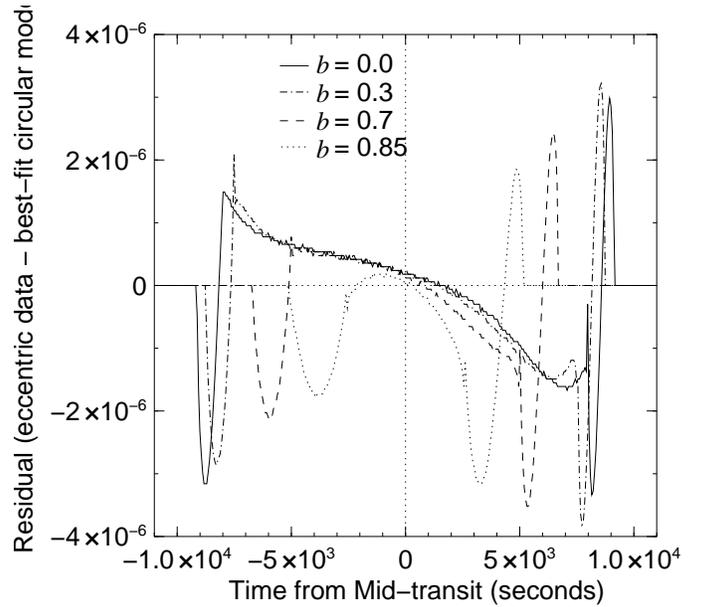}
\caption{Detectability of the transit lightcurve asymmetry induced by the orbital
eccentricity of planet HD147506b, for impact parameters $b=0.0$, $b=0.3$, $b=0.7$,
and $b=0.85$.
\label{figure:residual}}
\end{figure}

The degeneracy between $e$ and $f_0$, multiple cominations of which can produce
the same  $V_{f_0}$, can be broken by measuring the transit asymmetry outlined in
Section \ref{section:asymmetry}.  To measure the detectability of the asymmetry, I
produce a synthetic lightcurve for HD147506b ($r_\mathrm{p}=1.18\mathrm{R_{Jup}}$,
$R_*=1.8\mathrm{R_\odot}$, $a_\mathrm{p}=0.0685~\mathrm{AU}$, $e=0.507$,
$f_0=4.84$, and assuming \citet{2001ApJ...552..699B} limb darkening coefficient
$c_1=0.64$) that I fit using a model planet with a circular orbit.  The 
residuals, which I refer to as the detectability, are shown in Figure
\ref{figure:residual} for transits at several impact parameters.

When faced with an ingress and egress of differing duration, the best-fit
circular-orbit planet model splits the difference.  On ingress for HD147506b,
which transits before periastron ($f_0=-1.44$), the real planet is moving more
slowly than the model planet.  In the difference lightcurve, the real planet hits
the stellar limb \emph{before} the model, such that at mid-ingress the real and
model planets are in the same place.  So the real-minus-model
detectability initially trends negative, reaches zero at mid-ingress, and then
trends positive until the end of the planet's ingress.  The detectability varies
slowly between the end of ingress and the beginning of egress, resulting from the
planet covering stellar areas that have been slightly differently limb-darkened. 

The process is reversed on egress.  With the planet now moving more quickly than
the circular-orbit model, the model begins its egress early so that the real
planet will have caught up with it by mid-egress.  Hence the real-minus-model
detectability is again initially negative, reaches zero by mid-egress, and then
trends positive until the end of egress for both the real and model planets.

Despite the large $\Delta V$ for HD147506b, the magnitude of the transit
lightcurve asymmetry induced by orbital eccentricity is rather low (Figure
\ref{figure:residual}), peaking at only $3\times10^{-6}$ of the stellar flux. 
This low detectability is unmeasurable given current capabilities, and will
probably  prove challenging in the future as well given inherent stellar
variability.  The large stellar radius of HD147506 increases $\Delta V$, but also
decreases $R_\mathrm{p}/R_*$, diminishing the asymmetry effect.  Hence though
measurement of the asymmetry induced by an eccentric planet orbit can, along with
$V_f$, uniquely determine both $e$ and the longitude of periastron, the small
magnitude and duration of the effect are such that a measurement is unlikely to be
practical.

For transiting terrestrial-sized planets, determination of orbital eccentricity is
even more difficult.  If the ingress and egress are temporally resolved, that
duration along with the total transit duration together set the transit impact
parameter and $V_f$, providing a minimum $e$ constraint as per Equations
\ref{eq:eminperi} and \ref{eq:eminapo}.  For objects with such small transit
depths as terrestrial planets, though, it will be difficult to temporally resolve
the planets' ingress and egress.

Another way to constrain $b$ for terrestrial planets would be to use the effect of
stellar limb darkening.  \citet{2003ApJ...585.1038S} showed that minimizing limb
darkening more precisely delineates the end of a planet's ingress and the
beginning of its egress.  However strong limb darkening, if well-understood, can
provide a mechanism to ascertain a transit's impact parameter.  Similar to the
analysis used by \citet{2007ApJ...655..564K} to study HD209458b, fixing limb
darkening coefficients based on theoretical or previousely determined values could
suffice to constrain $b$, which, with a previously measured $R_*$, would then
determine $V_f$ and allow for constraints on $e$.

\section{CONCLUSIONS}
 
For a given orbital semimajor axis, extrasolar planets on eccentric orbits are
more likely to transit than planets on circular orbits by a factor of
$(1-e^2)^{-1}$.  As the space-based transit surveys \emph{CoRoT} and \emph{Kepler}
discover transiting planets that are far enough from their parent stars to have
avoided tidal circularization, more highly eccentric planets will be found
preferentially.  This bias is straightforward to remove if the eccentricity and
longitude of periastron are known, as they could be given follow-up radial
velocity observations.

The duration of a transit is a function of the planet's tangential velocity at
mid-transit, $V_{f_0}$.  For eccentric planets $V_{f_0}$ is greatest at periastron
and smallest at apoastron.  Hence if a planet transits near periastron the
duration is shorter than that of an equivalent planet in a circular orbit,
and similarly transits that occur when a planet is near apoastron is longer
than those of the equivalent circularly orbiting planet.  It would be useful to be
able to use this effect to determine the orbital eccentricity of discovered
transiting extrasolar planets, either before or without radial velocity follow-up.

If fitting the resulting lightcurve with a model planet on a circular orbit with
the known planetary period, a systematic error results in the determination of the
transit parameters if fitting for $R_*$, $R_\mathrm{p}$, $b$, and $c_1$ as done by
\citet{2001ApJ...552..699B} for HD209458b.  If instead the model system uses an
assumed stellar radius measured by different means, then the transit velocity
$V_{f_0}$ can be measured.  However without another, independent measurement of
either $e$ or the planet's longitude of periastron, knowledge of $V_{f_0}$ cannot
alone determine those parameters.  It can set a lower limit on a planet's orbital
eccentricity.

The difference in $V_f$ between a planet's ingress and egress that results from
the planet's orbital accelerations can resolve the $e$ / longitude of periastron
degeneracy.  This velocity differential $\Delta V$ introduces an asymmetry into
the transit lightcurve:  ingress is longer than egress before periastron, and
shorter after periastron.  My model fits to synthetic eccentric planet transit
lightcurves show that the detectability of this asymmetry is small, of order
$3\times10^{-6}$ for recently discovered eccentric transiting planet HD147506b. 
An effect that small is undetectable using present techniques.  As HD147506b is
nearly a model candidate for which to observe this effect, it is unlikely that
transit lightcurve asymmetry will prove useful for determining the orbital
parameters of transiting planets using photometry alone.

Determination of orbital eccentricity is of critical importance for evaluating the
habitability of terrestrial-sized transiting planets discovered by \emph{Kepler}. 
As these planets have masses too low for radial velocity measurements to detect,
our only constraints on $e$ for these planets will be photometric.  If the stellar
radius can be assumed from other, prior measurements, then it is possible to use
theoretical stellar limb-darkening coefficients within the \emph{Kepler} bandpass
to measure the transit impact parameter.  This measurement would then constrain
$V_{f_0}$ and allow a lower limit to be placed on $e$.

No techniques currently available will be able to uniquely measure the orbital
eccentricity of the terrestrial extrasolar planets that \emph{Kepler} will
discover.  The lower limits on eccentricity described above will allow for a
statistical exploration of the eccentricity distibution of terrestrial planets. 
That distribution will serve to constrain the formation and evolution of such
planets, as it has done for giant planets.  However, whether or not any particular
\emph{Kepler} planet is truly habitable will remain unknown until its orbital
eccentricity can be measured.

\acknowledgements

JWB acknowledges the support of the NASA Postdoctoral Program, administered for
NASA by Oak Ridge Associated Universities, and the support of NASA's \emph{Kepler}
mission for publication costs.

\newpage

\end{document}